\documentstyle[12pt]{article}

\title{Isospin violation and meson-exchange models of the nucleon-nucleon 
interaction\thanks{invited talk given at XIIIth International Seminar on
High Energy Physics Problems (ISHEPP 13), {\em Relativistic Nuclear
Physics and Quantum Chromodynamics}, held at the Joint Institute for
Nuclear Research, Dubna, Russia, September 2-7, 1996.}}
\author{Sidney A.\ Coon\\Department of Physics\\New Mexico State 
University\\Las Cruces, NM 88003}
\date{  }
\begin{document}

\maketitle

\begin{abstract}
Traditionally isospin violation (charge symmetry breaking and charge
dependence) has been modeled by the exchange of mesons between two
nucleons.      Sources of isospin
violation in these models include meson mass differences, meson mixing (of
pseudoscalar, vector, and axial-vector mesons) and isospin violating
couplings of mesons with nucleons. I will review the calculated results of this 
theoretical approach and the $NN$ data on isospin violation.
\end{abstract}

\baselineskip 0.6cm               
\pagebreak

\section{Experimental evidence for charge dependence and charge asymmetry}
	 
	\subsection*{A classification scheme for isospin violation}

Isospin violation in the nucleon-nucleon system is usually described in 
terms of classes of possible isospin operators for nucleons 1 and 2, 
although the classification scheme of Henley and Miller 
\cite{Millerrev1} applies to any baryon isodoublet system (such as $^3$H 
- $^3$He).  Here I review this scheme of four classes, beginning with 
the dominant class I isospin or charge independent interactions and 
ending
with the class IV interactions which are the weakest.  Class I forces 
have no isospin dependence (1) or their dependence is proportional to
$\mbox{\boldmath $\tau$}(1)\!\cdot\!\mbox{\boldmath $\tau$}(2)$.  Class 
II forces maintain charge symmetry but break charge independence and are 
proportional to $\tau_3(1)\tau_3(2)$.  They are characterized as 
proportional to an isotensor $\tau_3(1)\tau_3(2) - 
{\textstyle\frac{1}{3}}\mbox{\boldmath 
$\tau$}(1)\!\cdot\!\mbox{\boldmath $\tau$}(2)$.  Class III forces break 
charge symmetry (and therefore necessarily charge independence) and are 
symmetric under the interchange of particles 1 and 2.  Class III forces 
are proportional to ($\tau_3(1) +\tau_3(2)$).  They do not cause isospin 
mixing in the two-body system  because the third component of total 
isospin commutes with $I^2$.    Finally the class IV 
forces also break charge symmetry and charge independence and are anti 
symmetric under the interchange of particles 1 and 2.  These forces are 
proportional to ($\tau_3(1) - \tau_3(2)$) or to $ [\mbox{\boldmath 
$\tau$}(1)\!\times\!\mbox{\boldmath $\tau$}(2)]_3 $.  Class IV forces
connect states of $I=0$ and $I=1$.  For the two-body system these are 
the spin triplet $^3L_j$ and spin singlet $^1L_j$ states.  Classes II 
and IV forces vanish for $pp$ and $nn$ systems while class III forces 
vanish for the $np$ system.  Class IV forces have no effect on an $nn$ or 
$pp$ system but cause spin-dependent isospin mixing effects in the $np$ 
system.

	\subsection*{Class II and III: low energy $NN$ scattering parameters}
The best evidence for Class II and III forces lies in the low energy 
parameters of the effective range expansion of $np$ ($I=1$), $pp$, 
and $nn$ spin-singlet scattering and the $^3$H - $^3$He binding 
energy difference, 
after the effects of one-photon exchange have been removed.  
 I will utilize here the 
results of the Nijmegen phase shift analyses which include such small effects 
as vacuum polarization and the Breit (relativistic correction) term in 
one photon exchange.  The subtraction of one-photon exchange
to arrive at a pure nuclear effective range expansion from the recent 
Nijmegen phase-shift analyses is still underway, so I
will quote the low energy parameters of the recent strong $NN$
potentials \cite{Nij} which match the data nearly as well as the phase-shift
analysis. They are \cite{Stoks} 

\pagebreak

\begin{eqnarray*} 
    a_{np} \approx -24.0\;{\rm fm}  &  r_{np} \approx 2.68 \;{\rm fm} \\
    a_{pp} \approx -17.4\;{\rm fm}  &  r_{pp} \approx 2.83 \;{\rm fm} \\
\end{eqnarray*}
The experimental values of $a_{nn}$ found from $\pi^- d\rightarrow 
\gamma nn$ in which
only the photon was detected \cite{Gabioud} are
\begin{eqnarray*}
    a_{nn} = -18.5 \pm 0.4 \;{\rm fm}   & r_{nn} = 2.80 \pm
0.11 \;{\rm fm}  
\end{eqnarray*}
in excellent agreement with the kinematically complete determination of
$a_{nn} = -18.7 \pm 0.6$ fm from the same reaction \cite{Shori}. Other
compilations of the effective range parameters can be found in reviews
\cite{Millerrev2,Millerrev3}. There 
is a large difference between $np$ ($I=1$), and the average $pp$ 
and $nn$ $^1S_0$ scattering parameters ($\Delta a^{II} \equiv |a_{np}| - 
\textstyle\frac{1}{2}(|a_{nn}| + |a_{pp}|)$, $\Delta r^{II}_0 \equiv 
r_{np} - \textstyle\frac{1}{2}(r_{nn} + r_{pp})$)
\[   \Delta a^{II} \approx 6.0  \;\;{\rm fm} \hspace{1in} \Delta 
r^{II}_0 \approx -0.13 \;\;{\rm fm}  \]
indicating a fairly strong class II force (but still at the level of a 
few \% of the charge independent strong force).

The class III force effects on the effective range parameters 
($\Delta a^{III} \equiv |a_{nn}| - |a_{pp}|$, $\Delta r^{III}_0 \equiv
r_{nn} - r_{pp}$)
\[   \Delta a^{III} \approx 1.10  \;\;{\rm fm} \hspace{1in} \Delta
r^{II}_0 \approx -0.03 \;\;{\rm fm}  \]
are not so dramatic as are those of class II. They are, however, quite 
consistent 
in sign and magnitude with the positive value for the $^3H - ^3He$ 
binding energy difference of 764 keV, also due in part to class III 
nuclear forces.  The direct electromagnetic contribution to this number 
(static Coulomb force between the $pp$ pair in $^3He$ and other smaller 
electromagnetic effects such as vacuum polarization and the Breit 
interaction) must be subtracted to arrive at the effect from class III 
forces.  The totality of these effects was estimated first in a nearly 
model independent manner twenty years ago \cite{BCS} and that estimate 
has been verified by direct Faddeev calculations with many $NN$ 
potential models \cite{Wu} and by a combination of the two methods 
\cite{Bra88}.  Subtracting out these direct em contributions and a small 
difference in the $nn$ and $pp$ systems due to the neutron-proton mass 
difference, we find a remaining \cite{CN1996}
\[  \Delta E_{\rm exp}  \equiv  (^3H - ^3He)  \approx 76 \pm 24 
\;{\rm keV} \]
which is attributed to class III forces.

The positive $\Delta a$ reflects an
interaction between two neutrons which is more attractive than between
two protons and more binding energy is provided for $^3H$ as compared to
$^3He$.  The consistency in magnitude is more interesting.
It has recently been touted as a triumph of sophisticated Faddeev and 
quantum Monte Carlo calculations with modern $NN$ potentials that 
$\Delta E$ in the $A=3$ system can be explained by a class III charge 
asymmetric $NN$ force which has been adjusted to match 
$\Delta a^{III}$ and $\Delta r_0^{III}$ in the $A=2$ system 
\cite{Wu,Bra88,v18}.  However it has long been known from separable 
potential models that $\Delta E$ is much more sensitive to $\Delta 
r_0^{III}$ than to $\Delta a^{III}$ \cite{Karchenko}.  Gibson and 
Stephenson made a dedicated study of this dependence for both charge 
dependence and charge asymmetry in the bound trinucleon \cite{GS}.
For central separable potentials producing the correct $^3He$ binding 
their charge asymmetry results can be very well fitted by 
\begin{equation}
         \Delta E_{\rm GS} = ( 40 \Delta a -1600 \Delta r_0 )
\;{\rm keV/fm.} \label{eq:GS}
\end{equation}
(A discussion of the validity of this relationship in light of current 
three-body technology can be found in \cite{CN1996}).  The sensitivity 
of the trinucleon binding energy to $r_0$ rather than to $a$ was 
reviewed in the 1930's by Bethe and Bacher \cite{BB} with the model that 
Thomas used to demonstrate that nuclear forces are not contact forces 
but have a finite (i.e., non-zero) range \cite{thomas}.  In that model the 
binding energy of the deuteron is kept fixed by increasing the depth of 
the potential as the range is decreased, and the $nn$ interaction is 
neglected. Thomas demonstrated that, as the range of the two-body force 
goes to zero, the three-body binding energy becomes infinite.

	\subsection*{Class IV: $np$ elastic scattering}

The class II and class III effects in the $NN$ system require careful 
subtraction of much larger isospin violating one-photon exchange 
(including static Coulomb) to be 
revealed. For example, $a^{exp}_{pp} = -7.8014 \pm 0.0011$ fm has as a strong 
analogue $a_{pp} \sim 17.4$ fm after the subtraction \cite{Stoks}.  The 
large static Coulomb interaction is absent in $np$ elastic scattering so 
the small class IV charge asymmetry is not masked as is class III.
In a charge symmetric world the complete separation of isosinglet 
($I=0$) and isotriplet ($I=1$) states leads to a decoupling of spin 
singlet and spin triplet states and the equality of the analyzing powers 
when polarized neutrons are scattered from unpolarized protons and vice 
versa.  Any non-zero difference of the analyzing powers $\Delta A \equiv 
A_n - A_p$, where the subscripts denote the polarized nucleon, is 
evidence for the class IV forces which mix $I=0$ and $I=1$ states in the 
$np$ system.  Three (challenging) measurements have been made of a 
non-zero $\Delta A$ (at the zero-crossing angle of the average analyzing 
power) with increasing precision for the later experiments:
\begin{eqnarray*}
  \Delta A & = (47 \pm 22 \pm 8) \times 10^{-4}  
      & \;{\rm at\;477\;MeV}~\cite{triumf1} \\
   \Delta A & = (34.8 \pm 6.2 \pm 4.1) \times 10^{-4}
      & \;{\rm at\;183\;MeV}~\cite{IUCF}  \\
   \Delta A & = (59 \pm 7 \pm 2) \times 10^{-4}
      & \;{\rm at\;347\;MeV}~\cite{triumf2}\;. \\
\end{eqnarray*}
When one-photon exchange is removed from the third and latest 
measurement, leaving only the strong $NN$ interaction, $\Delta A$ is five 
standard deviations away from zero.

\section{Predictions of meson-exchange models}

A single or two-meson exchange graph is reduced to a non-relativistic 
potential which breaks charge independence or charge symmetry.  To 
calculate effects in  $NN$ systems the resulting potential is added to a 
charge independent phenomenological potential and the resulting change 
in the scattering parameters is compared to the experimental measures of 
section 1.

	\subsection*{Class II charge dependence}

The sum of class II forces from meson mass differences in single and 
two-meson exchange and $\gamma-\pi$ exchange does not explain the 
observed charge dependence.  For example, $\Delta a^{II}\sim 3.6$ fm 
from the sources discussed below falls short of the experimental value 
of $\sim 6$ fm.  This statement is contrary to previous conclusions 
because a recent calculation of the charge dependent $\gamma-\pi$ 
potential finds its effect to be an order of magnitude smaller than 
previous (flawed) estimates.

		\subsubsection*{Single meson exchange}

	Much of the observed class II effect in the low energy 
scattering parameters is accounted for by the mass difference of the 
charged and neutral pion.  A typical result is that found by Coon and 
Scadron \cite{Coonscad} for the pion mass difference ($\Delta m_{\pi} = 
m_{\pi^+} - m_{\pi^0}$) and rho mass difference (defined in the same 
way):
\begin{eqnarray*}
	\Delta a^{II} & \approx & (+2.9\; {\rm fm})_{\Delta m_{\pi}} + 
		(+0.1\;{\rm fm})_{\Delta m_{\rho}}  \\
	\Delta r_0 ^{II} & \approx & (-0.10\; {\rm fm})_{\Delta m_{\pi}} + 
		(0.0\;{\rm fm})_{\Delta m_{\rho}}  
\end{eqnarray*}
  These charge dependent 
potentials are due solely to the mass difference and the coupling 
constants are assumed to be charge independent.  The latter assumption 
is justified {\em a posteriori} by the Nijmegen phase shift analysis 
which finds almost no deviation from charge independence of the 
pion-nucleon coupling constant (Table 1 of \cite{vkfg}).  The poorly 
known vector meson mass difference is that given by the Coleman-Glashow 
tadpole model (to be discussed later).  In the above I quote results
obtained with a charge independent (CI) potential with a super-soft core
\cite{dtRS} which I think would be close to results with more modern CI
potentials.

		\subsubsection*{Simultaneous $\pi$ and $\gamma$ exchange}

Quite recently Friar and Coon reported the first reliable calculation of 
the isospin violating effect of simultaneous exchange of a pion and 
photon \cite{friarcoon}.  The calculation corresponds to a subset of 
leading order diagrams in chiral perturbation theory and one hopes that 
it corresponds to the dominant subset.  They worked in the static limit
($m_N \rightarrow \infty$) and in Coulomb gauge for the photon exchange.
The two Feynman diagrams which survive under those stipulations are than 
reduced to charge dependent potentials each of which gives a small 
$\Delta a^{II}$ of opposite sign.  The total effect is 
$\Delta a^{II}_{\gamma-\pi}\approx -0.15 \pm 0.03$ for a variety of short-range 
cutoffs and model CI potentials.  To the order worked out, the only 
effect of this mechanism is class II charge dependence and 
$\Delta a^{II}_{\gamma-\pi}$ is small (2-3\%) and of the opposite sign 
to the empirical $\Delta a^{II}$.  Because only the first terms in a 
$1/m_N$ expansion were kept, the class III or IV charge asymmetry is 
expected to be ${\cal O} (m_{\pi}/m_N)$ of the charge dependent result.

This class II $\gamma-\pi$ force result is about an order of magnitude 
smaller than previous estimates of $\Delta a^{II}_{\gamma-\pi} \approx +1$ 
fm which are quoted in the literature \cite{Millerrev2,Millerrev3,Ericson}.

		\subsubsection*{Meson mass difference in  $2\pi$ exchange}
 Ericson and Miller \cite{Ericson} employed the CI two-pion exchange 
potentials derived by Partovi and Lomon \cite{PL} from covariant Feynman 
graphs to estimate $\Delta a^{II}_{2\pi} \approx +0.9$ fm from $\Delta 
m_{\pi}$ in box and cross-box diagrams with (equal mass) nucleon 
intermediate states.  No effect due to the nucleon mass differences was
considered.  Coon and Scadron had earlier estimated the latter effect to 
be $\approx -0.15$ fm, also with the Partovi-Lomon potentials.  Chung and 
Machleidt \cite{CM} employed a Bonn potential to establish 
$\Delta a^{II}_{2\pi} \approx +0.9$ fm from nucleon and delta baryon 
intermediate states, only 0.2 of which came from nucleon intermediate 
states.  The initial polemics of this agreement 
\cite{Millerrev2,Machleidt} can be enjoyed in Refs. \cite{Ericson,CM}.

	\subsection*{Class III charge asymmetry in $\Delta a$, $\Delta 
r_o$, and $\Delta E$}

		\subsubsection*{$\Delta I$ = 1 meson mixing (single meson 
				exchange)}
The mixing of $I=0$ and $I=1$ mesons give rise to a class III $NN$ 
force \cite{PSC}.  This mixing occurs  in nature and is successfully
described by the Coleman-Glashow picture that the $\Delta I = 1$
hadronic processes are dominated by symmetry breaking  tree level
tadpole diagrams \cite{CG,Coleman}.  In this picture, at the quark
level, the meson-mixing matrix element 
$\langle a_1^\circ|H_{em}|f_1\rangle$ is determined by
the dominant single-quark operator
$H^{(3)} = {\textstyle \frac{1}{2}}(m_u - m_d)\bar{q}\lambda_3 q$,
established in Ref.~\cite{CS95} from the electromagnetic mass
differences in the pseudoscalar mesons, vector mesons, baryon octet, and
baryon decuplet.  When extended to the off-diagonal $\Delta I = 1$
transitions $\langle\pi^\circ|H_{em}|\eta_{NS}\rangle$, 
$\langle\rho^\circ|H_{em}|\omega\rangle$,   and 
$\langle a_1^\circ|H_{em}|f_{1\,NS}\rangle$ this gives the value
$-0.005$ GeV$^2$ for the mixing matrix elements, independent of the
particular mesons concerned.  Indeed, for this one-body operator, it is
to be expected that one obtains the same numerical value connecting any
$I=1$ state with an $I=0$ nonstrange state of the same spin and parity.
It also follows from this one-body operator structure (or the 
concomitant tadpole diagram at the hadronic level) that the $\Delta I = 
1$ transition has no dependence upon the four-momentum squared of the 
mesons, thus allowing it to be tested against experiment on-mass-shell 
for time-like $q^2$ {\em and} to be employed in the spacelike momentum 
transfer of an $NN$ force diagram.  These tests are passed for the 
vector $\rho\omega$ mixing \cite{CB} and pseudoscalar $\pi\eta\eta'$ 
mixing \cite{CMS86}, and, of course, there is no experimental data on 
the $\Delta I =1$ transitions of the axial vector mesons $a_1$ and 
$f_1$.

Explicit calculation shows \cite{CS82,CB,axial}
\begin{eqnarray*}
   \Delta a_{\pi\eta\eta'}^{III}  \approx  +0.26\;{\rm fm}  &  
r^{III}_o \approx -0.02 \;{\rm fm} & \Delta E \approx 32 \;{\rm keV}  \\
   \Delta a_{\rho\omega}^{III}  \approx  +1.5 \;{\rm fm}  &  
r^{III}_o \approx -0.03 \;{\rm fm} & \Delta E \approx 90 \;{\rm keV}  \\
   \Delta a_{a_1f_1}^{III}  \approx  +0.13 \;{\rm fm}  &  
r^{III}_o \approx =0.00 \;{\rm fm} & \Delta E \approx 15 \;{\rm keV}  \\
\end{eqnarray*}
indicating a good description of class III charge asymmetry by meson 
mixing alone.  The two-body results quoted are with the (charge 
independent) de 
Touriel-Sprung-Rouben potential and the three-body estimate was made 
with the ``model-independent" method \cite{BCS} also used to establish 
the empirical $\Delta E$.

		\subsubsection*{$2\pi$ exchange}

The major contribution to the difference of the $nn$ and $pp$ 
interactions comes from the baryon mass difference in intermediate 
states.  Unfortunately, a recent calculation based upon non-relativistic 
$\pi NN$ and $\pi N \Delta$ vertices \cite{CN1996} finds a much 
stronger effect ($\Delta a^{III} \approx +1.0\;{\rm fm}, \Delta r_0^{III} 
\approx -0.02\;{\rm fm}$, and $\Delta E \approx 43$ keV) than the first 
estimate \cite{Coonscad} which was based upon the Partovi-Lomon 
potential.  Neither the covariant calculation nor the non-relativistic 
calculation of the charge independent $2\pi$-exchange $NN$ potential 
have a clear chiral symmetric character.  Weinberg and van Kolck have
emphasized the utility of an explicit consideration of chiral symmetry in
the analysis of isospin violating interactions \cite{Wbg,Bira}.  Leading
order chiral $2\pi$-exchange potentials are now available \cite{chiralpot}
and could provide an alternative and more reliable foundation for future
studies of this mechanism.

A second source of class III (and class IV) forces, which is significant 
in describing the class IV experiments \cite{Jouni}, is baryon mass 
differences in non-relativistic $\pi NN$ and $\pi  N \Delta$ couplings 
themselves.  We found them to play a small role in class III effects:
($\Delta a^{III} \approx -0.1\;{\rm fm}, \; \Delta r_0^{III}
\approx 0.0\;{\rm fm}$, and $\Delta E \approx -1$ keV) \cite{CN1996}.
 
	\subsection*{Class IV charge asymmetry}
In meson-exchange based $NN$ potential models, the major strong 
interaction contribution to $\Delta A$ stem from single-pion exchange 
(due to the $np$ mass difference in the charged pion-$NN$ vertex just 
mentioned) and $\rho\omega$ mixing.  A class IV contribution from 
one-photon exchange (interaction of the proton current with the 
neutron magnetic moment) must be included in the analysis but it is 
comparatively small, especially at the higher energies of the TRIUMF 
measurements.  The $\rho\omega$ mixing contribution is similar to single 
photon exchange in that it is due to the product of the Pauli (magnetic) 
coupling of the isovector $\rho$ and the Dirac coupling of the isoscalar 
omega to the nucleons. (Mixing of pseudoscalar mesons cannot therefore 
give a class IV force.  Such a force could arise, in principle, from the
mixing of axial vector mesons but the Pauli couplings of axial vector
mesons to nucleons were neglected in Ref. \cite{axial} as being too
speculative at the present time.)

The low energy IUCF measurement cannot be described without $\rho\omega$ 
mixing,  although with slightly larger vector-nucleon couplings than used 
for the class III aspect of $\rho\omega$ mixing.  In any event, the 
theoretical predictions of Holzenkamp, Holinde, and Thomas \cite{HHT} and 
Iqbal and Niskanen \cite{IN} agree well with all 
three measurements (see Fig. 3 of Ref. \cite{triumf2}).

\section{Effective field theories of isospin violation}
Recently, following a suggestion by Weinberg, isospin
violation has been studied in the context of a general effective Lagrangian
which also displays chiral symmetry breaking \cite{Bira}. The first steps
have been taken in parameterizing this approach with the aid of the nuclear data
\cite{vKFriar}.  The effective field approach does not depend upon the
models discussed in section 2 but the conclusions of this first paper is
consistent with the results of the single-meson exchange models.  For lack
of space, I cannot discuss further this promising approach which links in a
direct way the observed isospin violation in the $NN$ interaction to the
symmetries of QCD.

The meson-exchange models, unlike the effective field theories, take their 
strengths and structure from hadronic physics outside the $NN$ interaction 
and therefore make predictions.  Although not directly related to QCD, 
they do,  however, provide a reasonable picture of 
class III charge symmetry breaking and are not as successful in 
quantitatively describing class II charge dependence.

\end{document}